\documentclass[conference]{IEEEtran}
\IEEEoverridecommandlockouts
\usepackage{cite}
\usepackage{amsmath,amssymb,amsfonts}
\usepackage{algorithmic}
\usepackage{graphicx}
\usepackage{textcomp}
\usepackage{xcolor}
\usepackage{graphicx}
\usepackage{subcaption}
\usepackage{multirow}
\usepackage{url}
\usepackage[T1]{fontenc}
\urldef{\appendix}\url{http://www.public.asu.edu/~ssarka18/appendix.pdf}

\usepackage{color}
\def\BibTeX{{\rm B\kern-.05em{\sc i\kern-.025em b}\kern-.08em
		T\kern-.1667em\lower.7ex\hbox{E}\kern-.125emX}}
\begin{document}
	
	\title{Predicting enterprise cyber incidents using social network analysis on the darkweb hacker forums\\
		\thanks{Some of the authors are supported through the AFOSR Young Investigator Program (YIP) grant FA9550-15-1-0159, ARO grant W911NF-15-1-0282, and the DoD Minerva program grant N00014-16-1-2015.}
	}
	
	\author{\IEEEauthorblockN{Soumajyoti Sarkar}
		\IEEEauthorblockA{\textit{Arizona State University} \\
			Tempe, USA \\
			ssarka18@asu.edu}
		\and
		\IEEEauthorblockN{Mohammad Almukaynizi}
		\IEEEauthorblockA{\textit{Arizona State University} \\
			Tempe, USA \\
			malmukay@asu.edu }
		\and
		\IEEEauthorblockN{Jana Shakarian}
		\IEEEauthorblockA{\textit{Cyber Reconnaissannce, Inc.} \\
			Tempe, USA \\
			jana@cyr3con.ai}
		\and
		\IEEEauthorblockN{Paulo Shakarian}
		\IEEEauthorblockA{\textit{Arizona State University} \\
			Tempe, USA \\
			shak@asu.edu}
		
	}
	
	\maketitle
	
	\begin{abstract}
	With rise in security breaches over the past few years, there has been an increasing need to mine insights from social media platforms to raise alerts of possible attacks in an attempt to defend conflict during competition. We use information from the darkweb forums by leveraging the reply network structure of user interactions  with the goal of predicting enterprise cyber attacks. We use a suite of social network features on top of supervised learning models  and validate them on a binary classification problem that attempts to predict whether there would be an attack on any given day for an organization. We conclude from our experiments using information from 53 forums in the darkweb over a span of 12 months to predict real world organization cyber attacks of 2 different security events that analyzing the path structure between groups of users is better than just studying network centralities like Pagerank or relying on the user posting statistics in the forums.
	\end{abstract}
	
	\maketitle

	\section{Introduction} 
	\label{sec:intro}
	With the recent data breaches such as those of Yahoo, Uber, Equifax\footnote{https://www.consumer.ftc.gov/blog/2017/09/equifax-data-breach-what-do, https://www.consumer.ftc.gov/blog/2016/09/yahoo-breach-watch} among several others that emphasize the increasing financial and social impact of cyber attacks, there has been an enormous requirement for technologies that could provide such organizations with prior alerts on such data breach possibilities. These breaches are a direct or indirect result of cyber, electronic, and information operations to infiltrate systems and infrastructure as well as gain unauthorized access to information, thus setting an example of conflict during competition. On the vulnerability front, the Risk Based Security's VulnDB database\footnote{https://www.riskbasedsecurity.com/2017/05/29-increase-in-vulnerabilities-already-disclosed-in-2017/} published a total of 4,837 vulnerabilities in a quarter of 2017, which was around 30\% higher than previous year. This motivates the need for extensive systems that can utilize vulnerability associated information from external sources to raise alerts on such cyber attacks. The darkweb is one such place on the internet where users can share information on software vulnerabilities and ways to exploit them \cite{b1, b17}. Surprisingly, it might be difficult to track the actual intention of those users, thus making it necessary to use data mining and learning to identify the discussions among the noise that could potentially  raise alerts on attacks on external enterprises.  In this paper, we leverage the information obtained from analyzing the reply network structure of discussions in the darkweb forums to understand the extent to which the darkweb information can be useful for predicting real world cyber attacks. 
	
	Most of the work on vulnerability discussions on trading exploits in the underground forums \cite{b10, b15, b16}  and related social media platforms like Twitter\cite{b2, b9, b17} have focused on two aspects: (1) analyzing vulnerabilities discussed or traded in the forums and the markets, thereby giving rise to the belief that the ``lifecycle of vulnerabilities" in these forums and marketplaces and their exploitation have significant impact on real world cyber attacks \cite{b15, b16} (2)  prioritizing or scoring  vulnerabilities using these social media platforms or binary file appearance logs of machines to predict the risk state of machines or systems \cite{b8, b13}. These two components have been used in silos and in this paper, we ignore the steps between vulnerability exploit analysis and the final task of real world cyber attack prediction by removing the preconceived notions used in earlier studies where vulnerability exploitation is considered a precursor towards attack prediction. We instead hypothesize on user interaction dynamics conceived through posts surrounding these vulnerabilities in these underground platforms to generate warnings for future attacks. We note that we \textit{do not} consider whether vulnerabilities have been exploited or not in these discussions since a lot of zero-day attacks \cite{b13} might occur before such vulnerabilities are even indexed and their gravity might lie hidden in discussions related to other associated vulnerabilities or some discussion on exploits. The premise on which this research is set up is based on the dynamics of all kinds of discussions in the darkweb forums, but we attempt to filter out the noise to mine important patterns by studying whether a piece of information gains traction within important communities.
	
	 To this end, the major contributions of this research investigation are as follows:
	
	\begin{itemize}
		\item We create a network mining technique using the directed reply network of users who participate in the darkweb forums, to extract a set of specialized users we term $experts$ whose posts with \textit{popular vulnerability mentions} gain attention from other users in a specific time frame. 
		\item Following this, we  generate several time series of features that capture the dynamics of interactions centered around these $experts$ across individual forums as well as general social network and forum posting statistics based feature time series.
		\item We use these time series features and train a supervised learning model based on logistic regression with attack labels for 2 different events from an organization to predict daily attacks. We obtain the best results with an F1 score of 0.53 on a feature that explores the path structure between $experts$ and other users compared to the random (without prior probabilities) F1 score of 0.37. Additionally, we find superior performance of features from discussions that involve vulnerability information  over network centralities and forum posting statistics.
	\end{itemize}
	
	The rest of the paper is organized as follows: we introduce several terms and the dataset related to the vulnerabilities and the darkweb in Section~\ref{sec:dataset}, the general framework for attack prediction including feature curation and learning models in Section~\ref{sec:feat_define}, and finally the experimental evaluations in Section~\ref{sec:exp}.

	\section{Background and Dataset}
	\label{sec:dataset}
	In this section, we describe the dataset used in our research to analyze the interaction patterns of the users in the Darkweb and the real world security incidents\footnote{We would often use the terms attacks/incidents/events interchangeably} data that we use as ground truth for the evaluation of our prediction models.
	
	\subsection{Enterprise-Relevant External Threats (GT)}
\begin{figure}[!t]
	\minipage{0.2\textwidth}
	\includegraphics[width=4.5cm, height=3cm]{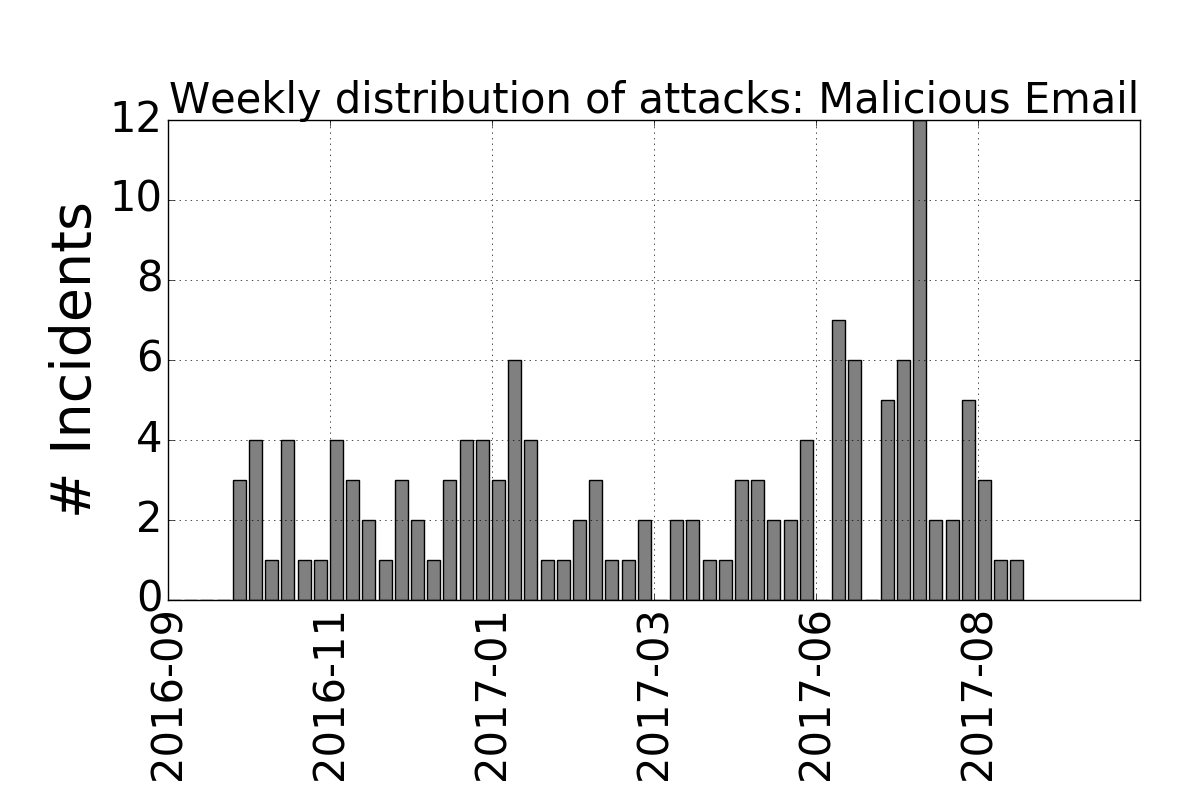}
	\subcaption{}
	\endminipage
	\hfill
	\minipage{0.25\textwidth}
	\includegraphics[width=4.5cm, height=3cm]{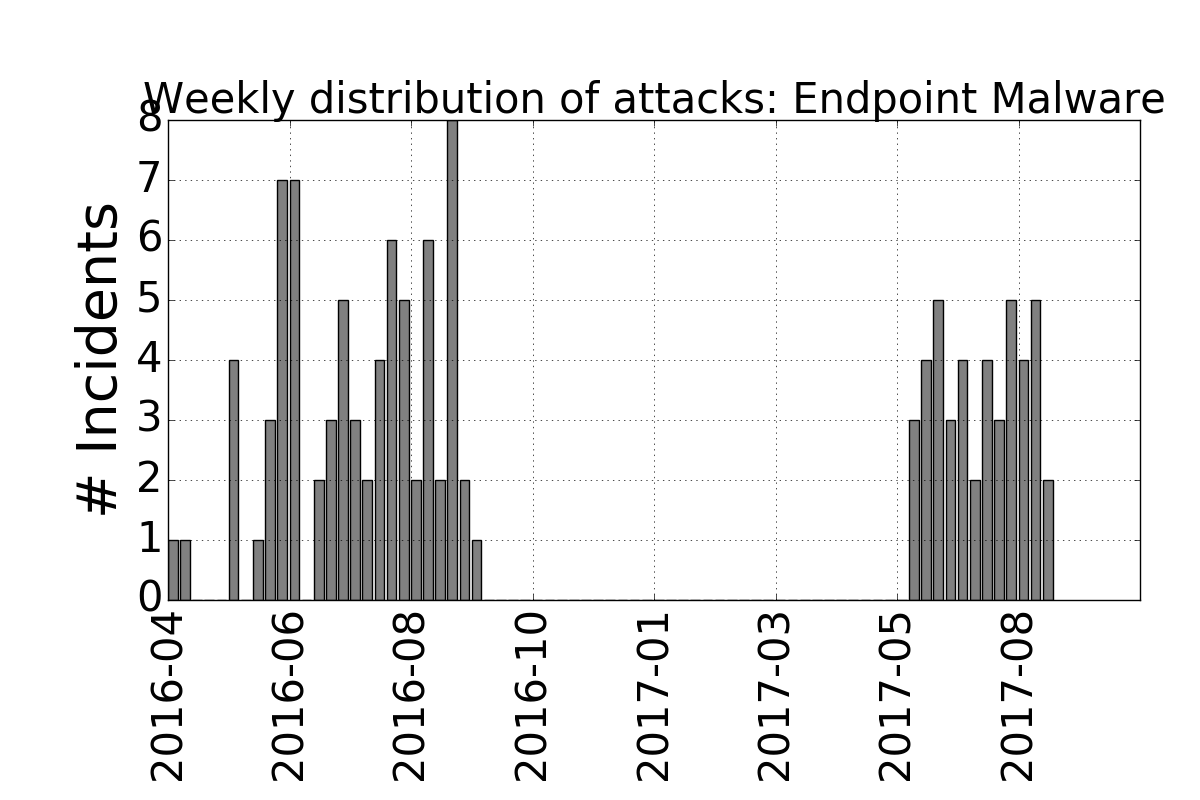}
	\subcaption{}
	\endminipage
	\hfill
	\caption{Weekly occurrence of security breach incidents of different types (a) Malicious email (b) Endpoint Malware}
		\label{fig:types_events}
	\end{figure}

	We use the Ground Truth (GT) available from the CAUSE program \footnote{https://www.iarpa.gov/index.php/research-programs/cause} that provided us with data from $Armstrong \ \ \ Corporation$  which contains information on cyber attacks on their systems in the period of April 2016 to September 2017. The data contains the following relevant attributes: \textbf{$\{$} \textit{event-type}: The type of attack called $event-type$ and \textit{event occurred date}: Date on which there was an attack of particular event-type. The \textit{event-types} that are used in this study are: \textit{Malicious email} refers to an event associated with an individual in the organization receiving an email that contains either a malicious attachment of link, and \textit{Endpoint Malware} refers to a malware on endpoint that is discovered on an endpoint device. This includes, but not limited to, ransomware, spyware, and adware. As shown in Figure~\ref{fig:types_events}, the distribution of attacks over time is different for the events. The total number of incidents reported for the events are as follows: 119 tagged as \textit{endpoint-malware} and 135 for \textit{malicious-email} events resulting in a total of 280 incidents over a span of 17 months that were considered in our study.
	
	\subsection{Darkweb data}
	The dark web forms a small part of the deep web, the part of the Web not indexed by web search engines, although sometimes the term deep web is mistakenly used to refer specifically to the dark web. We obtain all darkweb data used in this study through an API provided by a commercial platform\footnote{Data is provided by Cyber Reconaissance, Inc., www.cyr3con.ai}.
	
	A darkweb forum structure exhibits a hierarchical structure: each forum consists of several independent threads, a thread caters to a particular discussion on a topic, each thread spans several posts initiated by multiple users over time. We note that one user can appear multiple times in the sequence of posts depending on when and how many times the user posted in that thread. However the dataset we obtained does not contain the hierarchical information of reposting - it does not provide us with which user did a particular user, reply to, while posting or replying in a thread. We filter out forums based on a threshold number of posts that were created in the timeframe of January 2016 to September 2017. We gathered data from 179 forums in that time period where the total number of unique posts were 557,689 irrespective of the thread that they belonged to. The number of forums with less than 100 posts is large and therefore we only consider forums which have greater than 5,000 posts in that time period which gave us a total of 53 forums.  We denote the set of these 53 forums used in this dataset using the symbol $F$. \\
	\noindent \textbf{Common Vulnerabilities and Exposures (CVE):} The database of Common Vulnerabilities and Exposures maintained on a platform operated by the MITRE corporation \footnote{http://cve.mitre.org} provides an identity mapping for publicly known information-security vulnerabilities and exposures. We collect all the information regarding the vulnerability mentions in the darkweb forums in the period from January 2016 to October 2017. The total number of CVEs mentioned in the posts across all forums in this period are 3553.\\
	\noindent \textbf{CVE - CPE mapping: } A CPE  (Common Platform Enumeration) is a structured naming scheme for identifying and grouping clusters of information technology systems, software and packages maintained in a platform NVD (National Vulnerability Database) operated by NIST \footnote{https://nvd.nist.gov/cpe.cfm}. Each CVE can be assigned to different CPE groups based on the naming system of CPE families as described in \cite{b10}. Similarly, each CPE family can have several CVEs that conform to its vendors and products that the specific CPE caters to. In order to cluster the set of CVEs in our study into a set of CPE groups, we use the set of CPE tags for each CVE from the NVD database maintained by NIST. For the CPE tags, we only consider the operating system platform and the application environment tags for each unique CPE. Examples of CPE would include: \textit{Microsoft Windows\_95}, \textit{Canonical ubuntu\_linux}, \textit{Hp elitebook\_725\_g3}. The first component in each of these CPEs denote the operating system platform and the second component denotes the application environment and their versions. \\
	
	\section{Framework for Attack Prediction}\label{sec:feat_define}
	The mechanism for attack predictions can be described in 3 steps : (1) given a time point $t$ on which we need to predict an enetrprise attack of a particular event type (2) we use features from the darkweb forums prior to $t$ and, (3) we use these features as input to a learned model to predict attack on $t$. So one of the main tasks involves learning the attack prediction model, one for each event type.  Below we describe steps (2) and (3) - feature curation and building supervised learning models. 
	
	\subsection{Feature curation}
	
	We first describe the mechanism in which we build temporal networks following which we describe the features used for the prediction problem. We build 3 groups of features across forums: (1) Expert centric (2) User/Forum statistics (3) Network centralities.\\
	\noindent \textbf{Darkweb Reply Network}:
	We assume the absence of global user IDs across forums\footnote{Note that even in the presence of global user IDs across forums, a lot of anonymous or malicious users would create multiple profiles across forums and create multiple posts with different profiles, identifying and merging which is an active area of research.} and therefore analyze the social interactions using networks induced on specific forums instead of considering the global network across all forums. We denote the directed reply graph of a forum $f \in F$ by $G^f$ = $(V^f, E^f)$ where $V^f$ denotes the set of users who posted or replied in some thread in forum $f$ at some time in our considered time frame of data and $E^f$ denotes the set of 3-tuple $(u_1, u_2, rt)$ directed edges where $u_1, u_2 \in V^f$ and $rt$ denotes the time at which $u_1$ replied to a post of $u_2$ in some thread in $f$, $u_1 \rightarrow u_2$ denoting the edge direction. We denote by $G_{\tau}^f$ = $(V_{\tau}^f, E_{\tau}^f)$, a temporal subgraph of $G^f$, $\tau$ being a time window such that $V_{\tau}^f$ denotes the set of individuals who posted in $f$ in that window and $E_{\tau}^f$ denotes the set of tuples $(v_1, v_2, rt)$ such that $rt \in \tau$, $v_1, v_2 \in V_{\tau}^f$. We use 2 operations to create temporal networks: \textit{Create} - that takes a set of forum posts in $f$ within a time window $\tau$ as input and creates a temporal subgraph $G_\tau^f$  and \textit{Merge} - that takes two temporal graphs as input and merges them to form an auxiliary graph. To keep the notations simple, we would drop the symbol $f$ when we describe the operations for a specific forum in $F$ as context but which would apply for any forum $f$ $\in F$. We describe these two operations in brief, however a detailed algorithm relating the network construction is given in Algorithm 1 of  \textbf{Appendix A1}\footnote{Online Appendix: \appendix }. We adopt an incremental analysis approach by splitting the entire set of time points in our frame of study into a sequence of time windows $\Gamma$ = $\{\tau_1, \tau_2,  \ldots, \tau_{\mathcal{Q}} \}$, where each subsequence $\tau_i$, $i \in [1, \mathcal{Q}]$ is equal in time span and non-overlapping and the subsequences are ordered by their starting time points for their respective span. \\
	\begin{figure}[!t]
		\centering
		\includegraphics[width=8.5cm, height=3.6cm]{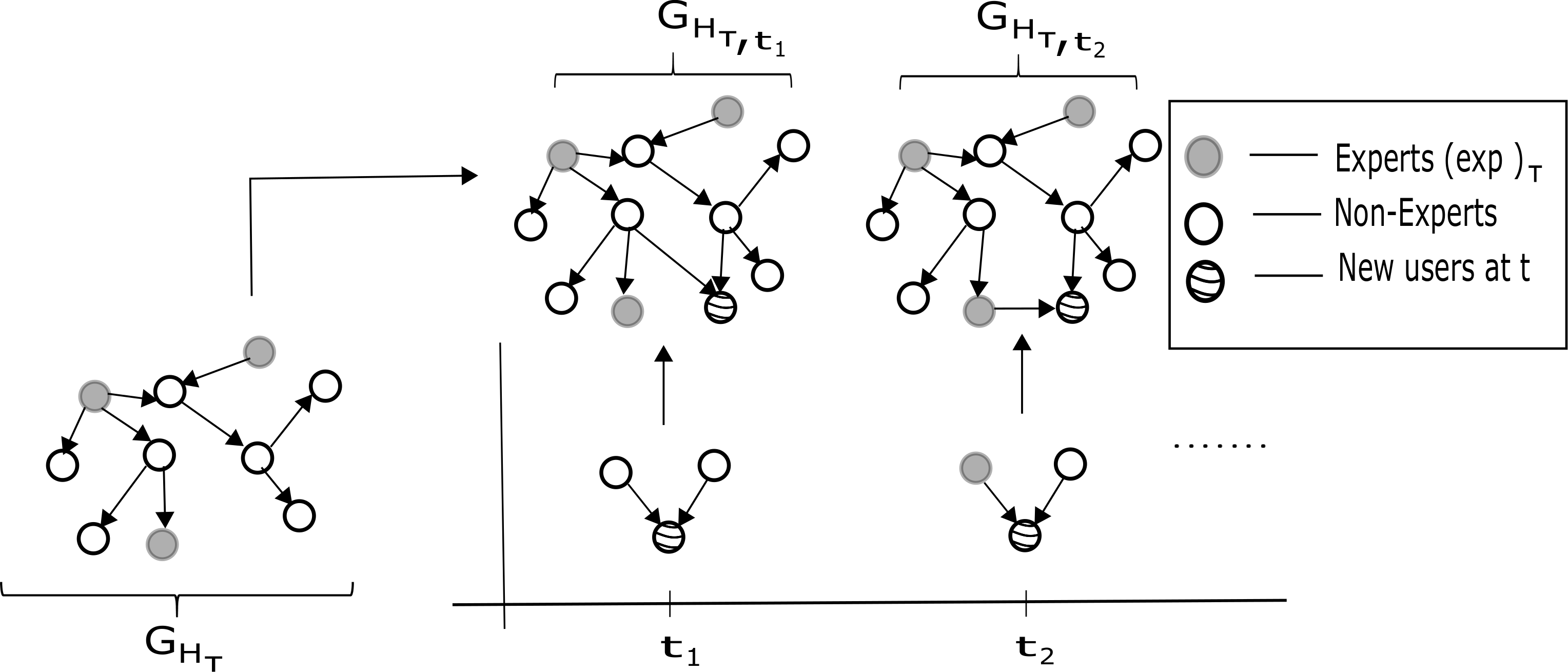}
		\caption{An illustration to show the \textit{Merge} operation: $G_{H_{\tau}}$ denotes the historical network using which the experts shown in gray are computed. $\{ G_{t_1}$, $G_{t_2}$, $\ldots \}$ denote the networks at time $t_1,  \ t_{2}, \ldots$ $\in \tau$, $\tau \in \Gamma$.}
		\label{fig:create_demo}
	\end{figure}
	\begin{table*}[!t]
		\begin{tabular}{|p{2.3cm}|p{2.7cm}|p{12cm}|}
			\hline
			Group                                 & Features                 & Description                                                                                                                              \\ \hline
			\multirow{4}{*}{Expert centric}       & Graph Conductance        & \begin{tabular}[c]{@{}l@{}}$\tau_{x}[t]$ = $\frac{\sum_{x \in exp_{\tau}} \sum_{y \in V_t \setminus exp_{\tau} } \pi(exp_{\tau}) P_{xy} }{\pi(exp_{\tau})}$ \\ where $\pi(.)$ is the stationary distribution  of the network $G_{H_\tau, t}$, $P_{xy}$ denotes the probability of \\ random walk  from vertices $x$ to $y$. The conductance represents the probability of taking a random walk \\ from any of the $experts$ to one of the users in $V_t \setminus exp_{\tau}$, normalized by the probability weight of \\ being on an expert. \end{tabular} \\ \cline{2-3} 
			& Shortest Path            & \begin{tabular}[c]{@{}l@{}}$\tau_{x}[t]$ = $ \frac{1}{|exp_{\tau}|} \  \sum_{e \in exp_{\tau} } \  \underset{u \in V_t \setminus exp_{\tau}}{\min} \  s_{e,u} $  \\ where $s_{e, u}$ denotes the shortest path from an expert $e$ to user $u$ following the direction of edges.\end{tabular}                                          \\ \cline{2-3} 
			& Expert replies           & \begin{tabular}[c]{@{}l@{}}$\tau_{x}[t]$ = $\frac{1}{|exp_{\tau}|} \  \sum_{e \in exp_{\tau}} \  |OutNeighbors(e)|$ \\ where $ OutNeighbors(.)$ denotes the out neighbors of user in the network $G_{H_\tau, t}$.\end{tabular}                                                                                                             \\ \cline{2-3} 
			& Common Communities       & \begin{tabular}[c]{@{}l@{}}$\tau_{x}[t]$ = $ \{ \mathcal{N}(c(u) \ | \ c(u) \in c_{experts} \wedge \ u \in V_t \setminus exp_{\tau} \}$\\ wher $c(u)$  denotes the community index of user $u$, $c_{experts}$ that of the experts and $\mathcal{N}(.)$ \\ denotes a counting function. It counts the number of users who share communities with experts. \end{tabular}                                      \\ \hline
			\multirow{4}{*}{Forum/User Statistcs}      & Number of threads        & $\tau_{x}[t]$ = $|\{ h \  | \  \mbox{thread h was posted on t} \}|$                                                                                                                                                                                                                                                                          \\ \cline{2-3} 
			& Number of users          & $\tau_{x}[t]$ = $|\{ u \  | \  \mbox{user u  posted on t} \}|$                                                                                                                                                                                                                                                                             \\ \cline{2-3} 
			& Number of expert threads & $\tau_{x}[t]$ = $|\{ h \  | \  \mbox{thread h was posted on t by users u $\in$ experts} \}|$                                                                                                                                                                                                                                                         \\ \cline{2-3} 
			& Number of CVE mentions   & $\tau_{x}[t]$ = $|\{ CVE \  | \  \mbox{CVE was mentioned in some post on t} \}|$                                                                                                                                                                                                                                                      \\ \hline
			\multirow{6}{*}{Network Centralities} & $Outdegree_{k}$          & $\tau_{x}[t]$ =  Average value of top k users, by outdegree on $t$                                                                                                                                                                                                                                                                    \\ \cline{2-3} 
			& $Outdegree_k$ CVE            & $\tau_{x}[t]$ =  Average value of top k users with more than 1 CVE mention in their posts, by outdegree on $t$                                                                                                                                                                                                \\ \cline{2-3} 
			& $Pagerank_k$                 & $\tau_{x}[t]$ =  Average value of top k users, by Pagerank on $t$                                                                                                                                                                                                                                                                    \\ \cline{2-3} 
			& $Pagerank_k$ CVE             & $\tau_{x}[t]$ =  Average value of top k users with more than 1 CVE mention in their posts, by pagerank  on $t$                                                                                                                                                                                                                                                                                               \\ \cline{2-3} 
			& $Betweenness_k$              & $\tau_{x}[t]$ =  Average value of top k users, by Betweenness on $t$                                                                                                                                                                                                                                                                    \\ \cline{2-3} 
			& $Betweenness_k$ CVE          & $\tau_{x}[t]$ =  Average value of top k users with more than 1 CVE mention in their posts, by betweenness on $t$                                                                                                                                                                                                                                                                                                 \\ \hline
		\end{tabular}
		\caption{List of features used for learning. Each feature $\tau_{x}$ is computed separately across forums. 	\label{tab:table_feat}}
	\end{table*}
	
	\noindent \textbf{CREATE:} \textit{ Creating the reply graph} - Let $h$ be a particular thread or topic within a forum $f$ containing posts by users $V_{h}^f$ = $\{u_1, \ldots, u_k \}$ posted at corresponding times $T_{h}^f$ = $\{t_1, \ldots , t_k\}$, where $k$ denotes the number of posts in that thread and $t_i \geq t_j $ for any $i > j$, that is the posts are chronologically ordered. To create the set of edges $E^f_{h}$, we connect 2 users $(u_i, u_j)$ $\in V_h^f$ such that $i > j$, that is user $u_i$ has potentially replied to $u_j$, and subject to a set of \textit{spatial} and \textit{temporal} constraints (\textbf{Appendix A1}).  These constraints make up for the absence of exact information about the reply hierarchies as to whom $u$ replied to in a particular post in $h$.\\
	\noindent \textbf{MERGE:} \textit{ Merging network} - In order to create a time series feature $\mathcal{T}_{x, f}$ for feature $x$ from threads in forum $f$ that maps each time point $t \in \tau$, $\tau \in \Gamma$ to a real number, we use 2 networks: (1) the historical network $G_{H_{\tau}}$ which spans over time $H_{\tau}$ such that $\forall t' \in H_{\tau}$, and $t \in \tau$, we have $t' < t$, and (2) the network $G^f_t$ induced by user interactions between users in $E_t$, which varies temporally for each $t \in \tau$. We note that the historical network $G_{H_{\tau}}$ would be different for each subsequence $\tau$ and same for all $t \in \tau$, so as the subsequences $\tau \in \Gamma$ progress with time, the historical network $G_{H_{\tau}}$ also changes, and we discuss the choice of spans $\tau \in \Gamma$ and $H_{\tau}$ in Section~\ref{sec:exp}. Finally, for computing feature values for each time point $t \in \tau$, we merge the 2 networks $G_{H_{\tau}}$ and $G_{t}$ to form the auxiliary network $G_{H_\tau, t}$ = $(V_{H_\tau, t}, E_{H_\tau, t})$, where $V_{H_\tau, t} = V_{H_\tau} \cup V_t$ and $E_{H_\tau, t}$ = $E_{H_\tau} \cup E_t$. A visual illustration of this method is shown in Figure~\ref{fig:create_demo}. Now we describe the several features we used that would be fed to a learning model for attack prediction. We compute time series of several features $x$, $\mathcal{T}_{x, f}[t]$ for every time point $t$ in our frame of study and for every forum $f$ separately.\\
	
	\noindent \textbf{1. Expert centric features}\\
	We extract a set of users we term $experts$ who have a history of CVE mentions in their posts and whose posts have gained attention in terms of replies. Following that, we mine several features that explain how attention is broadcast by these $experts$ to other posts. All these features are computed using the auxiliary networks $G_{H_{\tau}, t}$ for each time $t$. Our hypothesis is based on the premise that any unusual activity must spur attention from users who have knowledge about vulnerabilities. \\
	We focus on users whose posts in a forum contain most discussed CVEs belonging to \textit{important} CPEs over the timeframe of analysis, where the importance will shortly be formalized. For each forum $f$, we use the historical network $G_{H_{\tau}}^f$ to extract the set of $experts$ relevant to timeframe $\tau$, that is $exp_{\tau}^f$ $\in V_{H_{\tau}}^f$. First, we extract the top CPE groups $CP^{top}_{\tau}$ in the time frame $H_\tau$ based on the number of historical mentions of CVEs. We sort the CPE groups based on the sum of the CVE mentions in $\tau$ that belong to the respective CPE groups and take the top 5 CPE groups by sum in each $H_{\tau}$.  Using these notations, the experts $exp_\tau^f$ from history $H_{\tau}$ considered for time span $\tau$ are defined as users in $f$ with the following three constraints: (1) Users who have mentioned a CVE in their post in $H_\tau$. This ensures that the user engages in the forums with content that is relevant to vulnerabilities. (2) let $\theta(u)$ denote the set of CPE tags of the CVEs mentioned by user $u$ in his/her posts in $H_{\tau}$ and such that it follows the constraint: either $\theta(u)$ $\in CP_{\tau}^{top}$ where the user's CVEs are grouped in less than 5 CPEs or, $CP_{\tau}^{top}$ $\in \theta(u)$ in cases where a user has posts with CVEs in the span $H_{\tau}$, grouped in more than 5 CPEs.  This constraint filters out users who discuss vulnerabilities which are not among the top CPE groups in $H_\tau$ and (3) the in-degree of the user $u$ in $G_{H_\tau}$ should cross a threshold. This constraint ensures that there are a significant number of users who potentially responded to this user thus establishing $u$'s central position in the reply network. Essentially, these set of experts $exp_{\tau}$ from $H_{\tau}$ would be used for all the time points in $\tau$. We curate path and community based features based on these experts listed in Table~\ref{tab:table_feat}. These expert-centric features try to quantify the distance between an expert and a daily user(non-expert) in terms of how fast a post from that user receives attention from the expert. In that sense, the community features also measure the like-mindedness of non-experts and experts. \\
	\noindent \textbf{Why focus on experts?} To show the significance of these properties in comparison to other users, we perform the following hypothesis test: we collect the time periods of 3 widely known security events: the Wannacry ransomware attack that happened on May 12, 2017 and the vulnerability MS-17-010, the Petya cyber attack on 27 June, 2017 with the associated vulnerabilities CVE-2017-0144, CVE-2017-0145 and MS-17-010, the Equifax breach attack primarily on March 9, 2017 with vulnerability CVE-2017-5638. We consider two sets of users across all forums - $exp_{\tau}$, where $G_{H_\tau}$ denotes the corresponding historical network prior to $\tau$ in which these 3 events occurred and the second set of users being all $U_{alt}$ who are not experts and who fail either one of the two constraints: they have mentioned CVEs in their posts which do not belong to $CP^{top}$ or their in-degree in $G_{H_{\tau}}$ lies below the threshold.  We consider $G_{H_{\tau}}$ being induced by users in the last 3 weeks prior to the occurrence week of each event for both the cases, and we consider the total number of interactions ignoring the direction of reply of these users with other users. Let $\mathbf{deg_{exp}}$ denote the vector of count of interactions in which the \textit{experts} were involved and $\mathbf{deg_{alt}}$ denote the vector of counts of interactions in which the users in $U_{alt}$ were involved. We randomly pick number of users from $U_{alt}$ equal to the number of experts and sort the vectors by count. We conduct a 2 sample t-test on he vectors $\mathbf{deg_{exp}}$ and $\mathbf{deg_{alt}}$. The null hypothesis $H_0$ and the alternate hypothesis $H_1$ are defined as follows;
	$H_0: \ \mathbf{deg_{exp}} \leq \ \mathbf{deg_{alt}}$,  $H_1: \ \mathbf{deg_{exp}} > \ \mathbf{deg_{alt}}$. The null hypothesis is rejected at significance level $ \alpha $ = 0.01 with $p$-value of 0.0007. This suggests that with high probability, experts tend to interact more prior to important real world cybersecurity breaches than other users who randomly post CVEs. 
	
	Now, we conduct a second $t$-test where we randomly pick 4 weeks not in the weeks considered for the data breaches, to pick users $U_{alt}$ with the same constraints. We use the same hypotheses as above and when we perform statistical tests for significance, we find that the null hypothesis is not rejected at $\alpha$=0.01 with a $p$-value close to 0.05. This empirical evidence from the $t$-test also suggests that the interactions with $exp_{\tau}$ are more correlated with an important cybersecurity incident than the other users who post CVEs not in top CPE groups and therefore it is better to focus on users exhibiting our desired properties as experts for cyber attack prediction. Note that the $t-test$ evidence also incorporates a special temporal association since we collected events from three interleaved timeframes corresponding to the event dates.\\
	\noindent \textbf{2. User/Forum Statistics Features}\\
	We try to see whether the fourm or user posting statistics are themselves any indicators of future cyber attacks - for this we compute \textit{Forum/User Statistics} as described in Table~\ref{tab:table_feat}. 
	
	\noindent \textbf{3. Network centralities Features}\\
	In addition, we also tested several network \textit{Centrality} features mentioned in Table~\ref{tab:table_feat}. The purpose is to check whether emergence of central users in the reply network $G_{t}$, $t \in \tau$, are good predictors of cyber attacks. We note that in this case, we only use the daily reply networks to compute the features unlike the expert centric network features where we use $G_{H_\tau, t}$ .
	
	\subsection{Learning Models for Prediction} \label{sec:learn_models}
	
	In this section we explain how we use the time series data $\mathcal{T}_{x, f}$ to predict an attack at any given time point $t$. We consider a supervised learning model in which the time series  $\mathcal{T}_x$ is formed by averaging $\mathcal{T}_{x, f}$ across all forums in $f \in F$ at each time point $t$ and then using them for the prediction task. We treat the attack prediction problem in this paper as a binary classification problem in which the objective is to predict whether there would be an attack at a given time point $t$. Since the incident data in this paper contains the number of incidents that occurred at time point $t$, we assign a label of 1 for $t$ if there was at least one attack at $t$ and 0 otherwise. 

\begin{figure}[!t]
	\centering
	\includegraphics[width=4.5cm, height=1.5cm]{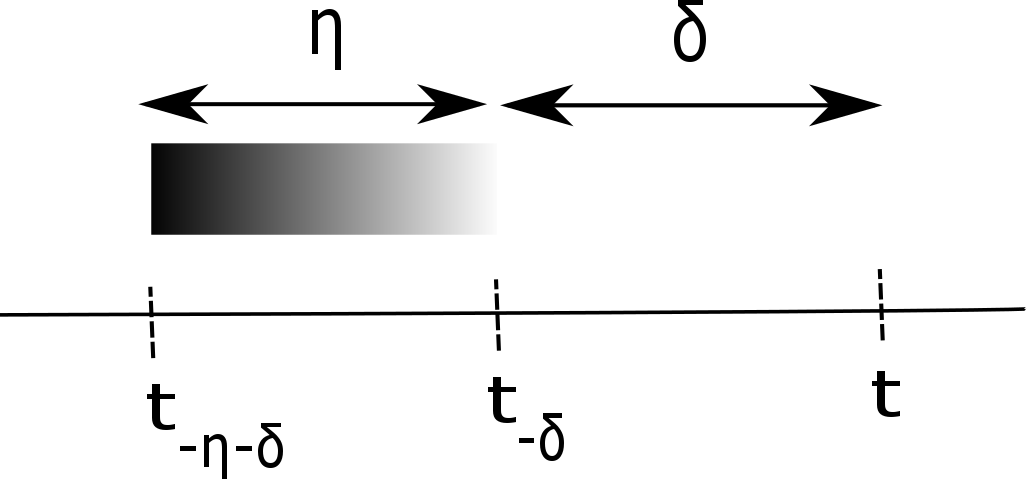}
	\caption{Temporal feature selection window for predicting an attack at time $t$}
	\label{fig:predict_des}
	\end{figure}

	In \cite{b4}, the authors studied the effect of longitudinal sparsity in high dimensional time series data, where they propose an approach to assign weights to the same features at different time spans to capture the temporal redundancy. 	We use 2 parameters: $\delta$ that denotes the start time prior to $t$ from where we consider the features for prediction and $\eta$, the time span for the features to be considered. An illustration is shown in Figure~\ref{fig:predict_des} where to predict an attack occurrence at time $t$, we use the features for each time $t_h$ $\in [t_{-\eta-\delta}, \  t_{-\delta}]$. Here we use logistic regression with longitudinal ridge sparsity that models the probability of an attack as follows with \textbf{X} being the set of features and $\mathbf{\beta}$ being the vector of coefficients:
	\begin{equation}
	P(attack(t) = \ 1 | \ \mathbf{X}  ) = \frac{1}{1+e^{-(\beta_0 + \sum_{k=\eta +\delta}^{\delta} \beta_k \ x_{t-k}) } }
	\end{equation}
	The final objective function to minimize over $N$ instances where $N$ here is the number of time points spanning the attack time frame is : $l(\mathbf{\beta}) = -\sum_{i=1}^{N} (y_i(\beta_0 + \mathbf{x_i}^T\mathbf{\beta} ) - \mbox{log}(1 + \mbox{exp}^{\beta_0 + \mathbf{x_i}^T\mathbf{\beta}}) +  \lambda \mathbf{\beta}^T\mathbf{\beta}$, $y$ being the instance label.
	
	 One of the major problems of the dataset is the imbalance in the training and test dataset as will be described in Section~\ref{sec:exp}, so in order to use all features in each group together for prediction, we use 3 additional regularization terms: the L1 penalty, the L2 penalty and the \textit{Group Lasso} regularization \cite{b5}. The final objective function can be written as:
	
	\begin{equation}\label{eq:3}
	l(\mathbf{\beta}) = -\sum_{i=1}^{N} \log(1 + e^{-y_i (\beta^T \mathbf{x_i}) }) + \frac{m}{2} \lVert \beta \rVert^2_2  + l\lVert \beta \rVert_1 + g.GL(\beta)
	\end{equation} 
	where $m$, $l$ and $g$ are the hyper-parameters for the regularization terms and the $GL(\beta)$ term is 
	$\sum_{g=1}^G \lVert \beta_{\mathcal{I}_g} \rVert_2$, where $\mathcal{I}_g$ is the index set belonging to the $g^{th}$ group of variables, $g = 1 \ldots G$. Here each $g$ is the time index $t_h$ $\in [t_{-\eta-\delta}, \  t_{-\delta}]$, so this group variable selection selects all features of one time in history while reducing some other time points to 0.  It has the attractive property that it does variable selection at the temporal group level and is invariant under (group-wise) orthogonal transformations like ridge regression. We note that while there are several other models that could be used for prediction that incorporates the temporal and sequential nature of the data like hidden markov models (HMM) and recurrent neural networks (RNN), the logit model allows us to transparently adjust to the sparsity of data, specially in he absence of a large dataset.

	\section{Experimental Evaluations} \label{sec:exp}
	In our work, the granularity for each time index in the $\mathcal{T}$ function is 1 day, that is we compute feature values over all days in the time frame of our study. For incrementally computing the values of the time series, we consider the time span of each subsequence $\tau \in \Gamma$ as 1 month, and for each $\tau$, we consider $H_\tau$ = 3 months immediately preceding $\tau$. That is, for every additional month of training or test data that is provided to the model, we use the preceding 3 months to create the historical network and compute the corresponding features on all days in $\tau$. For choosing the experts with an in-degree threshold, we select a threshold of 10 to filter out users having in-degree less than 10 in $G_{H_\tau}$ from $exp_{\tau}$. For the centralities features, we set $k$ to be 50, that is we choose the top 50 users sorted by that corresponding metric in Table~\ref{tab:table_feat}. We build different learning models using the ground truth available from separate $event-types$. 
	
	As mentioned in Section~\ref{sec:learn_models}, we consider a binary prediction problem in this paper - we assign an attack flag of 1 for at least 1 attack on each day and 0 otherwise have the following statistics: for \textit{malicious-email}, out of 335 days considered in the dataset, there have been reported attacks on 97 days which constitutes a positive class ratio of around 29\%, for \textit{endpoint-malware} the total number of attack days are 31 out of 306 days of considered span in the training dataset which constitutes a positive class ratio of around 26\%. For evaluating the performance of the models on the dataset, we split the time frame of each event into 70\%-30\% averaged to the nearest month separately for each $event-type$. That is we take the first 80\% of time in months as the training dataset and the rest 20\% in sequence for the test dataset. We avoid shuffle split as generally being done in cross-validation techniques in order to consider the consistency in using  sequential information when computing the features. As shown in Figures~\ref{fig:types_events}, since the period of attack information provided varies in time for each of the events, we use different time frames for the training model and the test sets. For the event \textit{malicious email} which remains our primary testbed evaluation event, we consider the time period from October 2016 to June 2017 (9 months) in the Darkweb forums for our training data and the period from July  2017 to August 2017 (3 months) as out test dataset, for the $endpoint-malware$, we use the time period from April 2016 to September 2016 (6 months) as our training time period and June 2017 to August 2017 (3 months) as our test data for evaluation.
	
		\begin{figure}
		\minipage{0.2\textwidth}
		\includegraphics[width=4cm, height=2.5cm]{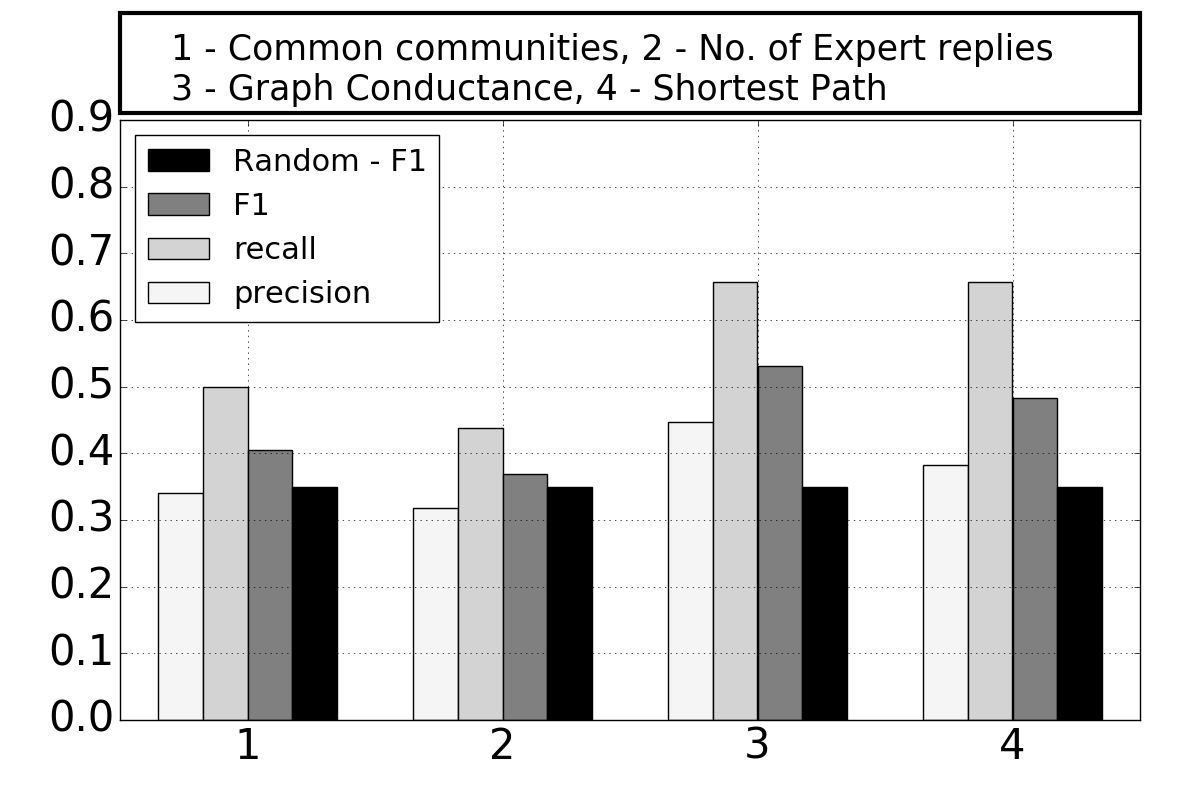}
		\subcaption{malicious-email}
		\endminipage
		\hfill
		\minipage{0.25\textwidth}
		\includegraphics[width=4cm, height=2.5cm]{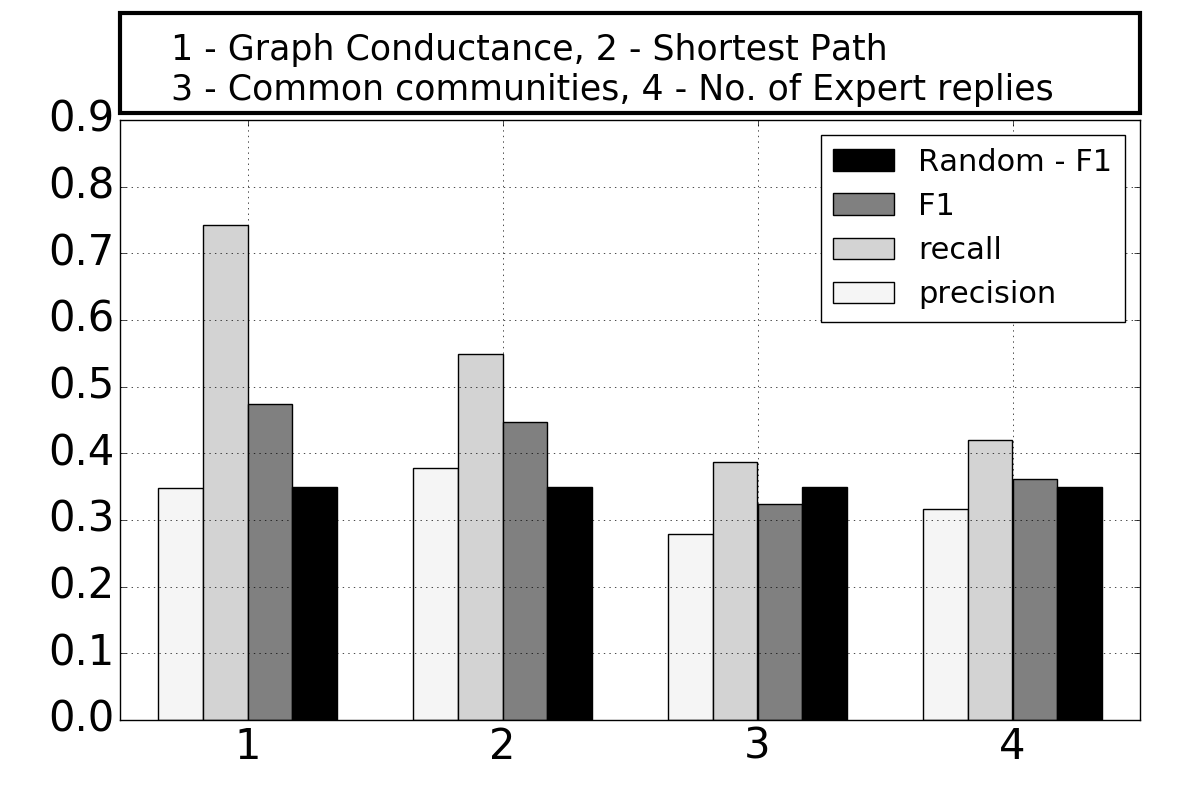}
		\subcaption{endpoint-malware}
		\endminipage
		\hfill
		\\
		\minipage{0.2\textwidth}
		\includegraphics[width=4cm, height=2.5cm]{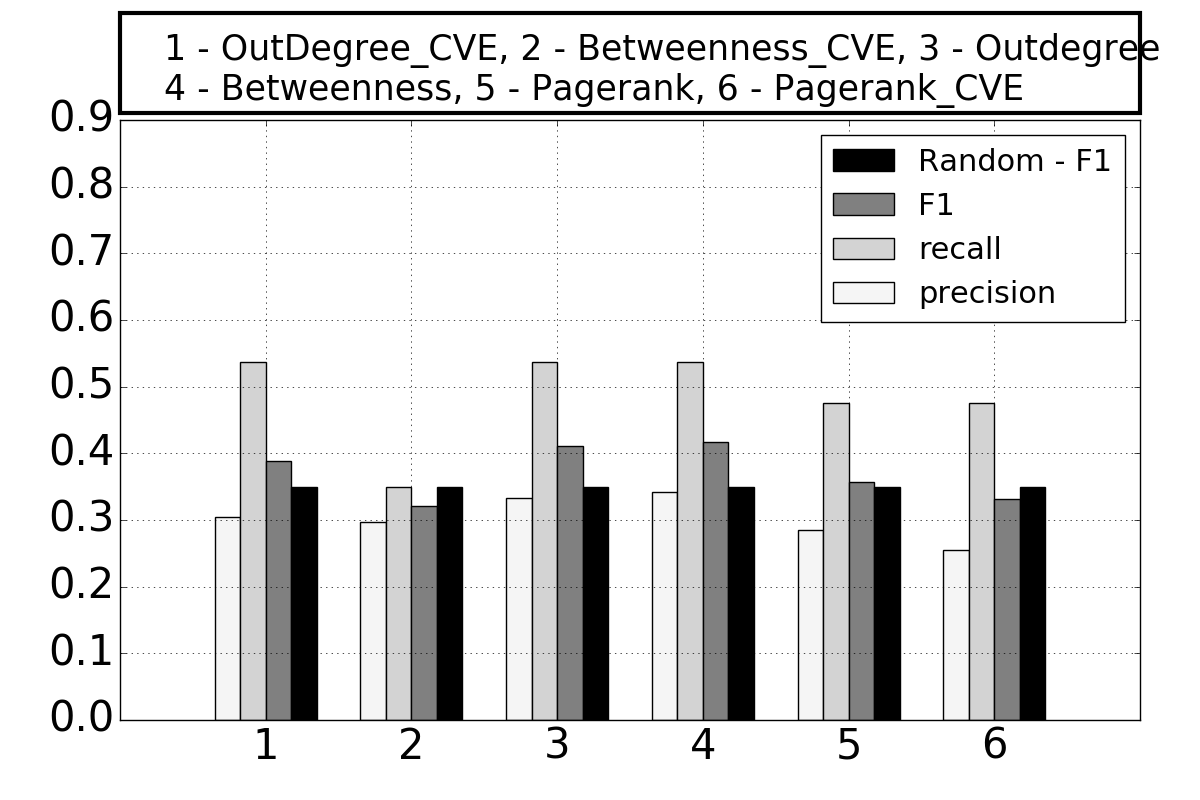}
		\subcaption{malicious-email}
		\endminipage
		\hfill
		\minipage{0.25\textwidth}
		\includegraphics[width=4cm, height=2.5cm]{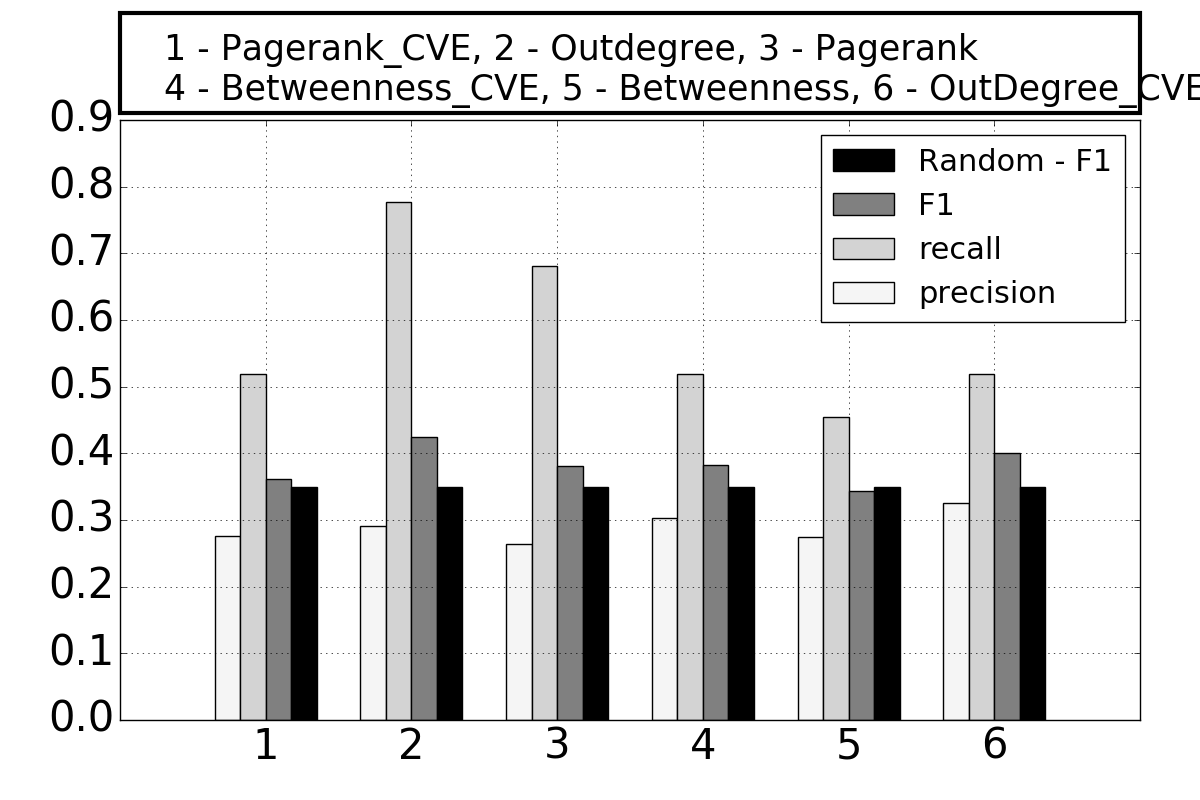}
		\subcaption{endpoint-malware}
		\endminipage
		\hfill
		\\
		\minipage{0.2\textwidth}
		\includegraphics[width=4cm, height=2.5cm]{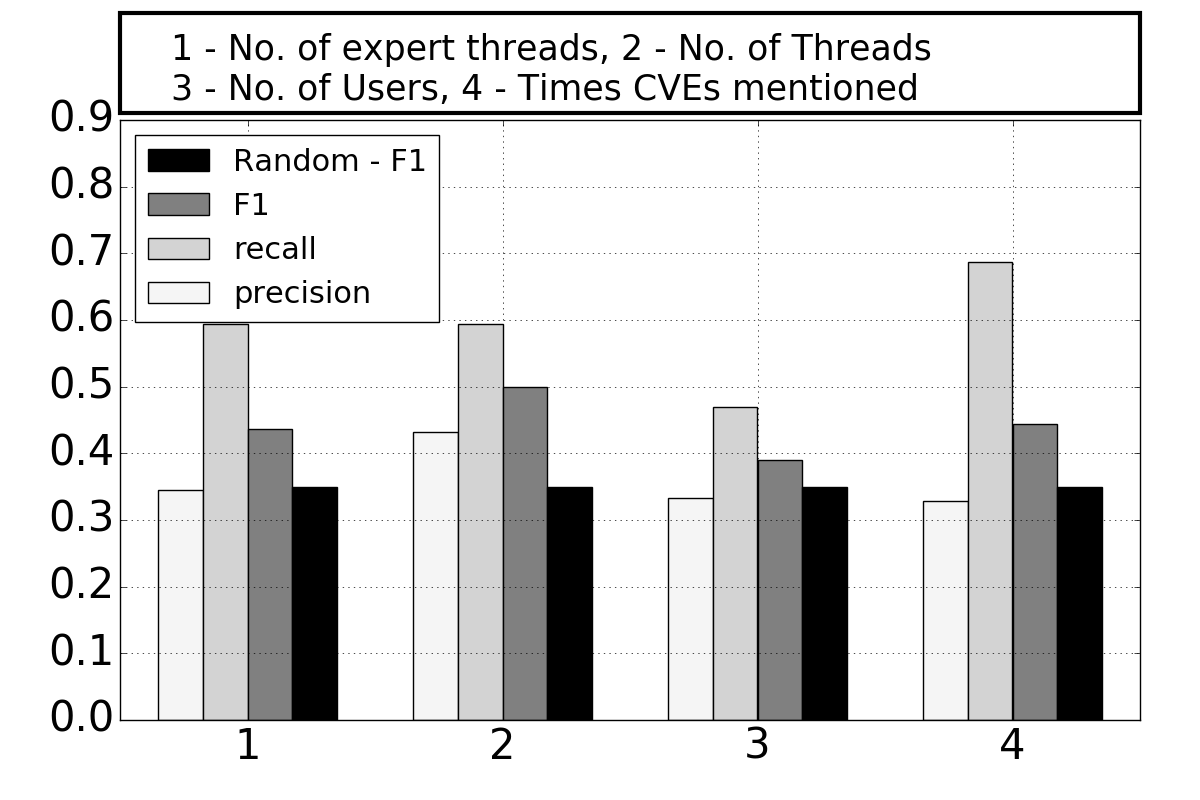}
		\subcaption{malicious-email}
		\endminipage
		\hfill
		\minipage{0.25\textwidth}
		\includegraphics[width=4cm, height=2.5cm]{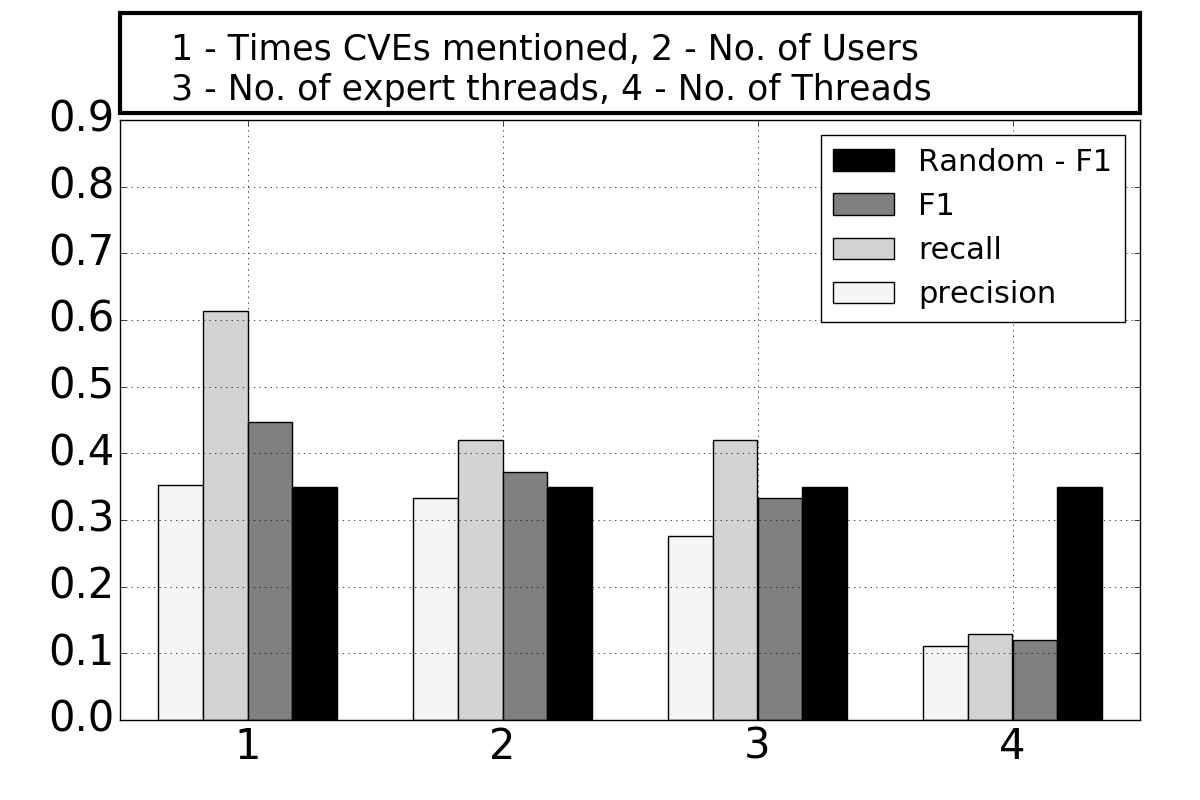}
		\subcaption{endpoint-malware}
		\endminipage
		\hfill
		\caption{Classification results for the features considering the logistic regression model: $\delta$ = 7 days, $\eta=8$ days.}
		\label{fig:class_7}
	\end{figure}

	We consider a span of 1 week time window $\eta$ while keeping $\delta$ = 8 days. From among the set of statistics features that were used for predicting $malicious-email$ attacks shown in Figure~\ref{fig:class_7}(e), we observe the best results using the number of threads as the signal for which we observe a precision of 0.43, recall of 0.59 and an F1 score of 0.5 against the random F1 of 0.34 for this type of attacks. From among the set of expert-centric features in Figure~\ref{fig:class_7}(a), we obtain the best results from graph conductance with a precision of 0.44, recall of 0.65 and an F1 score of 0.53 which shows an increase in recall over the number of threads measure. Additionally, we observe that  the best features in terms of F1 score are graph conductance and shortest paths whereas number of threads and vulnerability mentions turn out to be the best among the statistics. For the attacks belonging to the type $endpoint-malware$, we observe similar characteristics for the expert-centric features in Figure~\ref{fig:class_7}(b) where we obtain a best precision of 0.34, recall of 0.74 and an F1 score of 0.47 against a random F1 of 0.35, followed by the shortest paths measure. However for the statistics measures we obtain a precision of 0.35, recall 0.61 and an F1 score of 0.45 for the vulnerability mentions followed by the number of threads which gives us an F1 score of 0.43. Although the common communities features doesn't help much in the overall prediction results, in the following section we describe a special case that demonstrates the predictive power of the community structure in networks. On the other hand, when we investigate the centralities features with respect to the prediction performance in Figure~\ref{fig:class_7}(c), we find that just looking at network centralities does not help. The best values we obtain for \textit{malicious-email} event predictions are from the outdegree and betweenness metrics both of which gives us an F1 score of 0.41. Surprisingly, we find that when the metrics are used for only the users with CVE mentions, the results are worse with the best F1 score for outdegree CVE having an F1 score of 0.38. This calls for more complex understanding of path structures between users than just focusing on user significance solely. The challenging nature of the supervised prediction problem is not just due to the issue of class imbalance, but also the lack of large samples in the dataset which if present, could have been used for sampling purposes. As an experiment, we also used Random Forests as the classification model, but we did not observe any significant improvements in the results over the random case.
	
	For the model with the Group lasso regularization in Equation~\ref{eq:3}, we set the parameters $m, l, g$ and 0.3, 0.3 and 0.1 respectively. We obtain better results for each group of features together on the \textit{malicious-email} events with an F1 score of 0.55 for Expert centric, 0.51 with Forum/user statistics and 0.49 with network centrality based features. 
	
	\subsection*{Prediction in High Activity Weeks} 
	One of the main challenges in predicting external threats without any method to correlate them with external data sources like darkweb or any other database is that it is difficult to validate which kinds of attacks are most correlated with these data sources. To this end, we examine a controlled experiment setup for the $malicious-email$ attacks in which we only consider the weeks which exhibited high frequency of attacks compared to the overall timeframe: in our case we consider weeks having more than 5 attacks in test time frame. These high numbers may be due to multiple attacks in one or few specific days or few attacks on all days. We run the same supervised prediction method but evaluate them only on these specific weeks.
	
	\begin{figure}[!h]
		\centering
		\minipage{0.2\textwidth}
		\includegraphics[width=3.8cm, height=2.3cm]{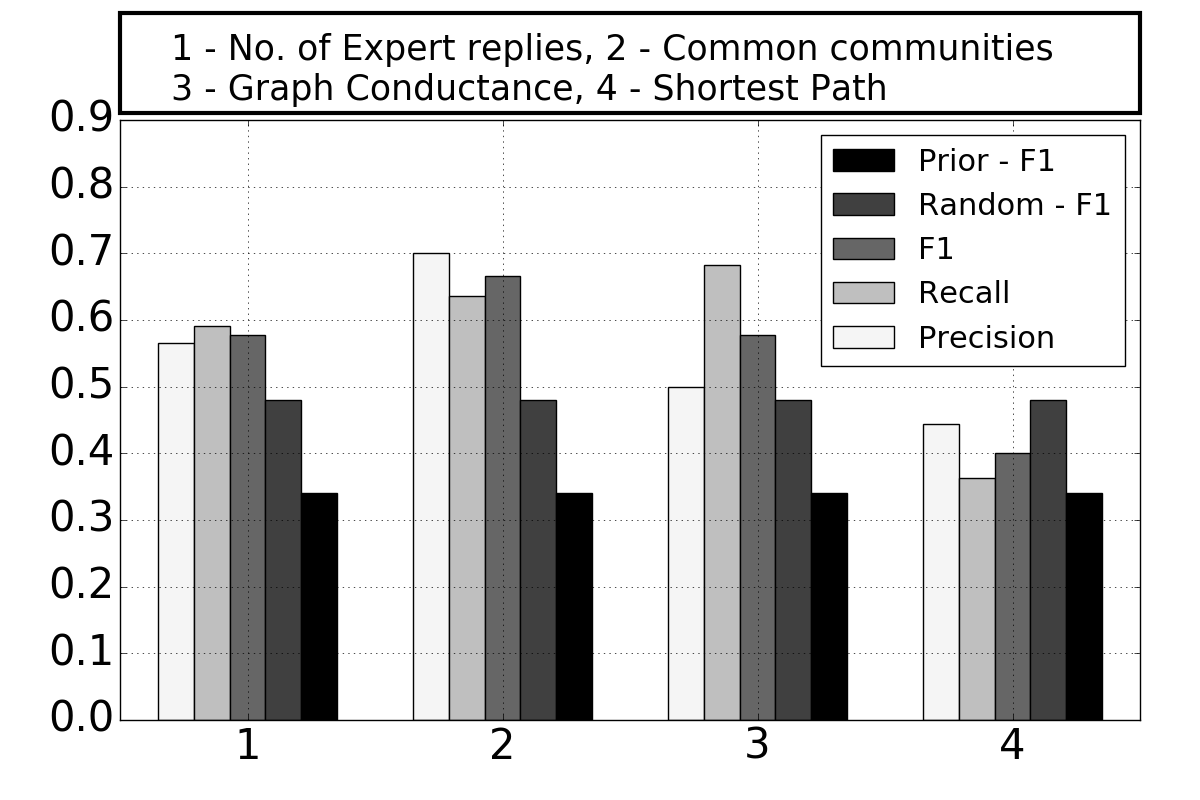}
		\subcaption{Expert centric}
		\endminipage
		\hfill
		\minipage{0.2\textwidth}
		\includegraphics[width=3.8cm, height=2.3cm]{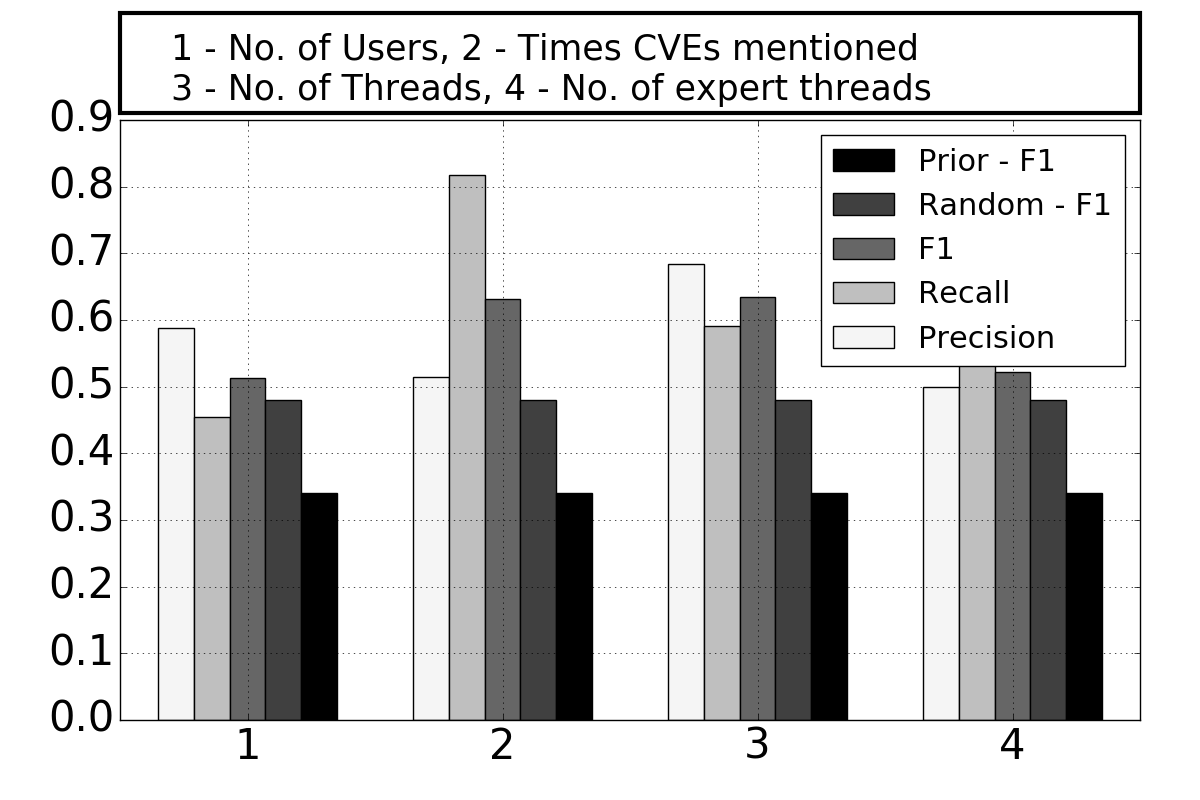}
		\subcaption{User/Forum Statistics}
		\endminipage
		\hfill
		\\
		\minipage{0.2\textwidth}
		\includegraphics[width=4cm, height=2.5cm]{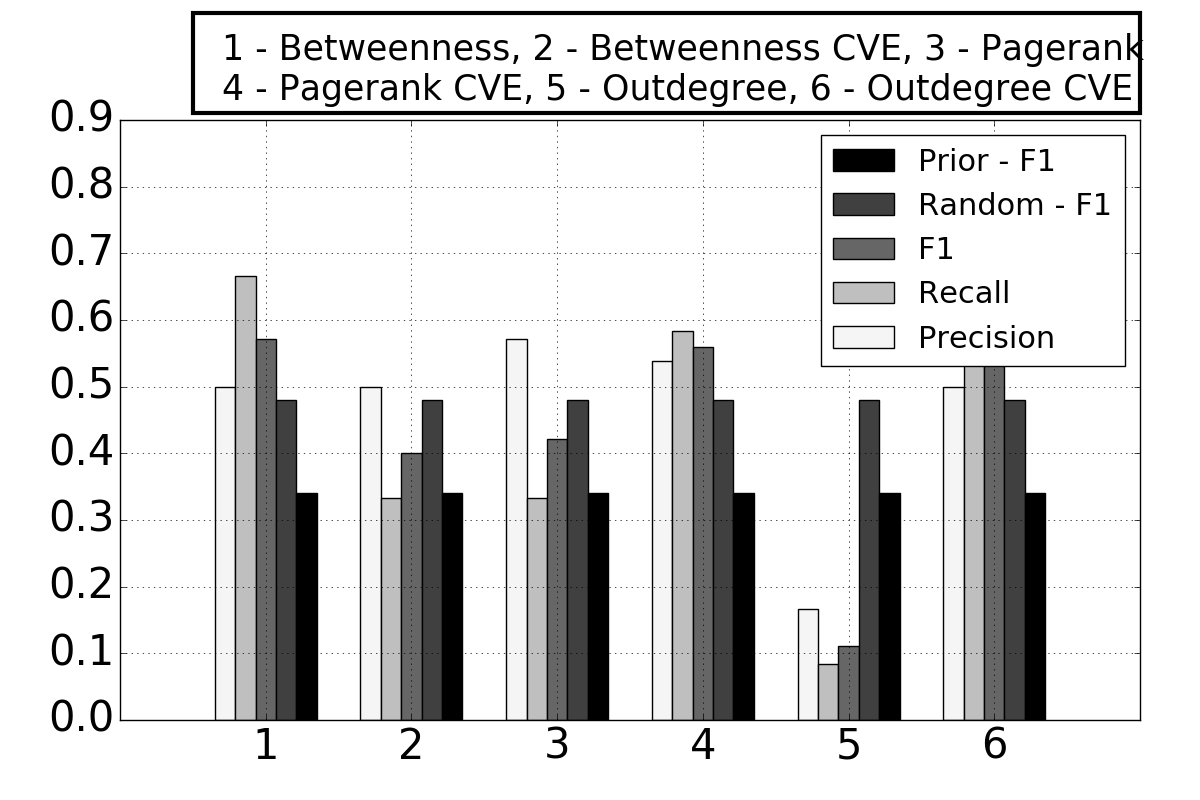}
		\subcaption{Centralities}
		\endminipage
		\hfill
		\caption{Classification results for $malicious-email$ attacks in high frequency weeks, $\delta$ = 7 days and $\eta$ = 8 days.}
		\label{fig:class_best}
	\end{figure}
	
	From the results shown in Figure~\ref{fig:class_best}, we find that the best results were shown by the common communities feature having a precision of 0.7 and a recall of 0.63 and an F1 score of 0.67 compared to the random (no priors) F1 score of 0.48 and a random (with priors) F1 score of 0.34 for the same time parameters. Among the statistics measures,  we obtained a highest F1 score of 0.63 for the vulnerability mentions feature. From among the set of centralities features, we find that betweenness measure has the best F1 score of 0.58 with a precision of 0.5 and a recall of 0.78. This also suggests the fact that analyzing the path structure between nodes is useful since betweenness relies on the paths passing through a node. Additionally, we find unlike the results over all the days, for these specific weeks, the model achieves high precision while maintaining comparable recall emphasizing the fact that the number of false positives are also reduced during these periods. This correlation between the weeks that exhibit huge attacks and the prediction results imply that the network structure analytics can definitely help generate alerts for cyber attacks.
	
	\section{Related Work and Conclusion}
	Using network analysis to understand the topology of Darkweb forums has been studied at breadth in \cite{b7} where the authors use social network analysis techniques on the reply networks of forums. There have been several attempts to use external social media data sources to predict real world cyber attacks \cite{b2, b9, b8}. Using machine learning models to predict security threats \cite{b2} has many open research fields including predicting whether a vulnerability would be exploited based on Darkweb sources \cite{b10, b3}. The availability of large external data sources makes the case for using machine learning methods for cyber attack prediction more promising. Previous studies also include using time series models for forecasting the number of cyber incidents \cite{b18} which motivates the need of such models for cyber attack prediction. The authors in \cite{b19} look at text mining techniques to understand the content of the posts in various social media platforms that provide threat intelligence. In this study, we argue that the darkweb can be a reliable source of information for predicting external enterprise threats. We leverage the network and interaction patterns in the forums to understand the extent to which they can be used as useful indicators. Our study also opens further research possibilities surrounding sentiment analysis on these discussions that could help  track the malicious discussions and hence defend against cyber conflict during competition.

\end{document}